\documentstyle[11pt,Sprocl,epsf,psfig]{article}

\def\be{\begin{equation}}
\def\ee{\end{equation}}
\def\bea{\begin{eqnarray}}
\def\eea{\end{eqnarray}}

\newcommand{\QrhoNote}{\,\footnote{A spheroidal system with
shortest-to-longest axis ratio $c/a \; (q_{\rho})$ of the density
contours has a shape E$n$, with $n$ such that $q_{\rho}=1-n/10$}}
\newcommand{\VLAnote}{\,\footnote{The VLA of the
National Radio Astronomy Observatory is a facility of the National
Science Foundation operated under cooperative agreement by Associated
Universities, Inc.} \ }
\newcommand{\etal}{\mbox{{\it et al. }}}
\newcommand{\qrho}{\mbox{$q_{\rho}$}}
\newcommand{\pmt}{\mbox{$\pm \;$}}
\newcommand{\rhoN}{\mbox{$\rho_0$}}
\newcommand{\Rc}{\mbox{$R_{\rm c}$}}
\newcommand{\hR}{\mbox{$h_{\rm R}$}}
\newcommand{\Kz}[1]{\mbox{$K_{\rm z}{#1}$}}
\newcommand{\kms}{\mbox{${\rm km \;s}^{-1}$}}
\newcommand{\Ht}{\mbox{${\rm H_2} \;$}}
\newcommand{\HI}{\mbox{{\rm H \footnotesize{I} }}}
\newcommand{\RSUN}{\mbox{$R_0 \;$}}
\newcommand{\VSUN}{\mbox{$\Theta_0 \;$}}
\newcommand{\MoverL}[1]{\mbox{$M/{\cal L}_{{\rm #1}} \;$}}
\newcommand{\Msun}{\mbox{$M_{\odot}$}}
\newcommand{\rtp}[1]{\mbox{$^{#1}$}}

\newcommand{\add}{\mbox{$^{\rm o} \!\!.$}}

\def\la{\mathrel{\hbox{\rlap{\hbox{\lower4pt\hbox{$\sim$}}}\hbox{$<$}}}}
\def\ga{\mathrel{\hbox{\rlap{\hbox{\lower4pt\hbox{$\sim$}}}\hbox{$>$}}}}

\def\Jour#1#2#3#4{{#1} {\bf #2}, #3 (#4)}
\def\Napj{{\em Astrophys. J.}}
\def\Napjl{{\em Astrophys. J. Letter}}

\def\Naj{{\em Astr. J.}}
\def\Nmnras{{\em Mon. Not. R. astr. Soc.}}
\def\Naap{{\em Astr. Astrophys.}}

\begin{document}

\title{The Flattened Dark Matter Halos of NGC 4244 and the Milky
Way\,\footnote{To appear in the proceedings of ``Aspects of Dark Matter
in Astro- and Particle Physics'', Heidelberg, September 1996 (World
Scientific; Eds. H.V. Klapdor-Kleingrothaus, Y. Ramachers).}}

\author{Rob P. Olling}
\address{University of Southampton, Department of
Physics and Astronomy, Southampton SO17 1BJ, United Kingdom, \\
E-mail: olling@astro.soton.ac.uk}

\maketitle

\abstracts{ In a previous paper\,\cite{PaperI} a method was developed to
determine the shapes of dark matter halos of spiral galaxies from the
flaring and velocity dispersion of the gas layer.  Here I present the
results for the almost edge-on Scd galaxy NGC
4244\,\cite{PaperII,PaperIII} and preliminary results for the Milky Way. 
NGC 4244's dark matter halo is found to be highly flattened with a
shortest-to-longest axis ratio of $0.2_{-0.1}^{+0.3}$.  If the dark 
matter is disk-like, the vertical velocity dispersion of the dark matter
must be $\sim$ 20\% larger than the measured tangential dispersion in
the \HI.  The flaring of the Milky Way's gas layer, the local column of
identified stars and the total column within 1.1 kpc from the plane are
consistent with a moderately flattened dark halo (E7-E0) and galactic
constants of \RSUN = 7.1 kpc, \VSUN = 180 \kms, while the stellar disk
is $\sim$ 80\% of maximal. 
}

\section{Introduction}

Although rotation curves of spiral galaxies have been used as evidence
for the presence of dark matter (DM), little is known about the nature,
extent and actual distribution of the DM in individual
galaxies\,\cite{ABBS85,LF89}.  The equatorial rotation curve, which probes
the potential in only one direction, provides no information about the
shape of DM halos. 
 
Several methods have been used to determine the shapes of dark matter
halos.  The analysis of warps\,\cite{HS94} shows that only one of the
five systems studied requires a DM halo as flattened as E4\QrhoNote.  On
the other hand, in studies of polar ring
galaxies\,\cite{SS90,SRJF94,SP95} substantially flattened DM halos are
found (E6-E7 and E5).  The shape of the dark halo of the Milky Way has
been estimated from stellar kinematics (E0 -
E7)\,\cite{BMO87,SZ90,vdM91,AC94}, and the method of flaring gas layers
(E0-E7; see below).  The S0 galaxy NGC 4753 seems to have a rather round
DM halo\,\cite{SKD92} (E1).  The X-ray isophotes of two early type
galaxies indicate that their dark halos are moderately
flattened\,\cite{BC96a} (E5.5 and E6).  Cold dark matter galaxy
formation simulations which include gas dynamics tend to produce rather
oblate DM halos\,\cite{jD94}, with $\qrho=c/a=0.5$ \pmt 0.15 (Fig. 1).

\vspace*{-5mm}
\psfig{file=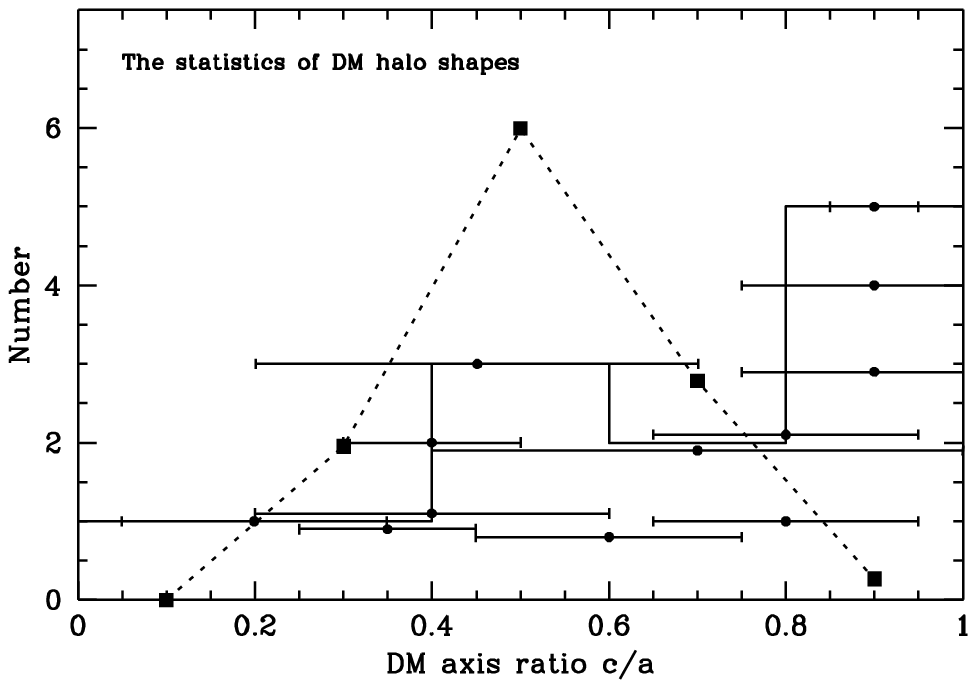,height=8cm,width=8cm}

\vspace*{-7.4cm}
\noindent
\parbox{3.7cm}{ Fig.  1. A histogram of the known DM halo shapes.  The
Dotted line and the filled squares represent the theoretical prediction. 
The points with error bars represent the individual galaxies.  Note the
discrepancy between the results from the warping-gas-layer method
(rightmost bin) and the other methods.}

\vspace*{3mm}

The shape of the dark halo can be determined by comparing the measured
thickness of the gas layer (flaring) with that expected from a
self-gravitating gaseous disk in the potential due to the stars and the
flattened DM halo\,\cite{PaperI,PaperIII}.  If a round halo, with a
certain density distribution, is squeezed along the vertical axis, the
densities and the exerted gravitational forces will increase, resulting
in a thinner \HI disk and higher rotation speeds.  In order to fit the
observed rotation curve, one has to deform the DM-halo such that the
DM-halo density ($\rho_{DM}$) at large distances will be roughly
inversely proportional to the halo flattening \qrho.  Since the
thickness of the gas layer beyond the optical disk is proportional to
$1/\sqrt{\rho_{DM}}$, \qrho $\propto$ (width of the gas layer)$^2$.

Below I apply the method to the galaxy NGC 4244 for which the basic
parameters were determined in Paper II\,\cite{PaperII}, and present
preliminary results for the Milky Way.

\section{The Method}

For a constant temperature gas, the density at height $z$ above the
plane, $\rho_{\rm gas}(z)$, can be calculated from the equation of
hydrostatic equilibrium:

\vspace*{-5mm}
\begin{eqnarray}
\sigma_{\rm gas}^2 \; \frac{d \; \ln{\rho_{\rm gas}(z)} }{d \; z}  &=&
  -\Kz{(z)} \; \; ,
\end{eqnarray}
\vspace*{-4mm}

\noindent with $\sigma_{\rm gas}$ the velocity dispersion.  Assuming
that the density distribution $\rho(z)$ extends to infinity (i.e.,
$\rho(R,\theta,z)=\rho(z)$) the potential has only a vertical gradient
so that the vertical force can be calculated from the Poisson equation:
$\Kz(z) = -4 \pi G \int_0^z \rho(z') dz'$.  Although not perfect, this
plane parallel sheet approximation has been used extensively in the past
and can be used for first order approximations to the expected thickness
of the gas layer\,\cite{PaperI}.  For example, in the cases that the
gaseous self-gravity, the stellar disk, or the DM halo dominates the
potential the full width at half maximum ($FWHM$) of the gas layer is
given by:

\vspace*{-5mm}
\begin{eqnarray}
FWHM_{gas}^{self}(R) &\approx& 1.58 \; \sigma_{gas}/\Sigma_{gas}
     \\
FWHM_{gas}^{stars}(R) &\approx& 0.6 \sqrt{z_e / \Sigma_{stars}} \;
   \times  \sigma_{gas}
     \\
FWHM_{gas}^{halo}(R) &\approx& 2.35
     \sqrt{ \left( \frac{2.4 \qrho}{1.4\; +\; \qrho} \right) } \;
     \left( \frac{\sigma_{gas}}{V_{max}} 
     \right) \sqrt{ R_c(\qrho)^2 + R^2 }
\end{eqnarray}
\vspace*{-4mm}

\noindent with $\Sigma_{gas}$ and $\Sigma_{stars}$ the gaseous and
stellar surface densities in units of \Msun pc\rtp{-2}, $z_e$ the
scale-height of the stellar disk, $V_{max}$ the maximum rotation speed
(in \kms), $R_c(\qrho)$ the core radius of the dark halo, and $R$ the
cylindrical radial distance (all distances in kpc).  These equations can
be combined to yield an approximate value for $FWHM_{gas}$ if more
than one component contributes to the potential\,\cite{PaperI}.  In
regions where the rotation curve rises steeply or the surface
density varies rapidly this approximation fails so that it
is better to calculate the vertical force from the density distribution
of the {\em whole} galaxy\,\cite{PaperI}:

\vspace*{-5mm}
\begin{eqnarray}
 \Kz{(R,z)} &=& G 
   \int_0^{\infty} r dr \rho(r,0) \int_{-\infty}^{\infty} dw \rho_{tot}(r,w)
   \int_{-\pi}^{\pi} \frac{d}{dz} 
   \frac{d\theta}{|\overline{\bf s} -\overline{\bf S}|} \; \; ,
\end{eqnarray}
\vspace*{-4mm}

\noindent with $\overline{\bf s}=\{r,\theta,w\}$, $\overline{\bf
S}=\{R,0,z\}$.  I incorporate three components in the global mass model;
1) a stellar disk with a density distribution which decreases
exponentially with radius and height above the plane; 2) a non-singular
flattened isothermal DM-halo with core radius \Rc(\qrho) and central
density \rhoN(\qrho) with a density distribution proportional to $1/R^2$
to reproduce the observed ``flat'' rotation curves; and 3) a gaseous
disk. 

Of course, the true DM-halo density distribution may be
different\,\cite{NFW96}.  However, for roundish DM distributions the
vertical force is roughly proportional to the radial force which is the
same for all disk-halo combinations that reproduce the observed rotation
curve, so that at large radii the width of the gas layer is independent
of the radial distribution of the DM. The flattening of the DM
halo introduces a $\sim \sqrt{\qrho}$-dependence on the thickness of the
gas layer. 

Comparing the thickness of the gas layer beyond the optical disk with
model flaring curves, calculated for a series of models with varying
halo flattening, then yields the halo shape. 

\section{Results: NGC 4244}

The Scd galaxy NGC 4244 was observed for 14 hours with the VLA\VLAnote
to determine the gaseous velocity dispersion and the rotation curve. 
While compact fast rotating galaxies are known to have declining
rotation curves\,\cite{CvG91,PSS96}, NGC 4244 is the only low mass
galaxy $(V_{max} \approx 100 \ \kms)$ for which the rotation curve
falls, in {\em Keplerian} fashion. 

A new technique\,\cite{PaperII} has been used to determine the width of
the gas layer (Fig.  2).  An upper limit to $FWHM_{gas}$ is found by
assuming a constant inclination of 84\add5, while incorporating the
slight warp into the analysis yields the best values for the flaring
(filled triangles). 

Comparing the observations with the model flaring curves (drawn lines), I
conclude that the DM halo of NGC 4244 is highly
flattened\,\cite{PaperIII}: \qrho = 0.2 \pmt 0.1 .  This flattening, the
most extreme value reported to date, lies at the extreme end of the
theoretical predictions (Fig.  1).  Is it possible that some systematic
effect plays a role and that NGC 4244's DM halo is less flattened? I
investigate several possibilities: a) The systematics introduced by the
uncertainty in the inclination is discussed in the caption of Fig.  1. 
b) Non-thermal pressure gradients due to magnetic fields and cosmic ray
heating [which add to the LHS.  of Eqn (1)] are unlikely to be important
beyond the optical disk because cosmic rays are closely related to sites
of star formation\,\cite{BH90}. If they {\em are} 

\vspace*{-14mm}
\psfig{file=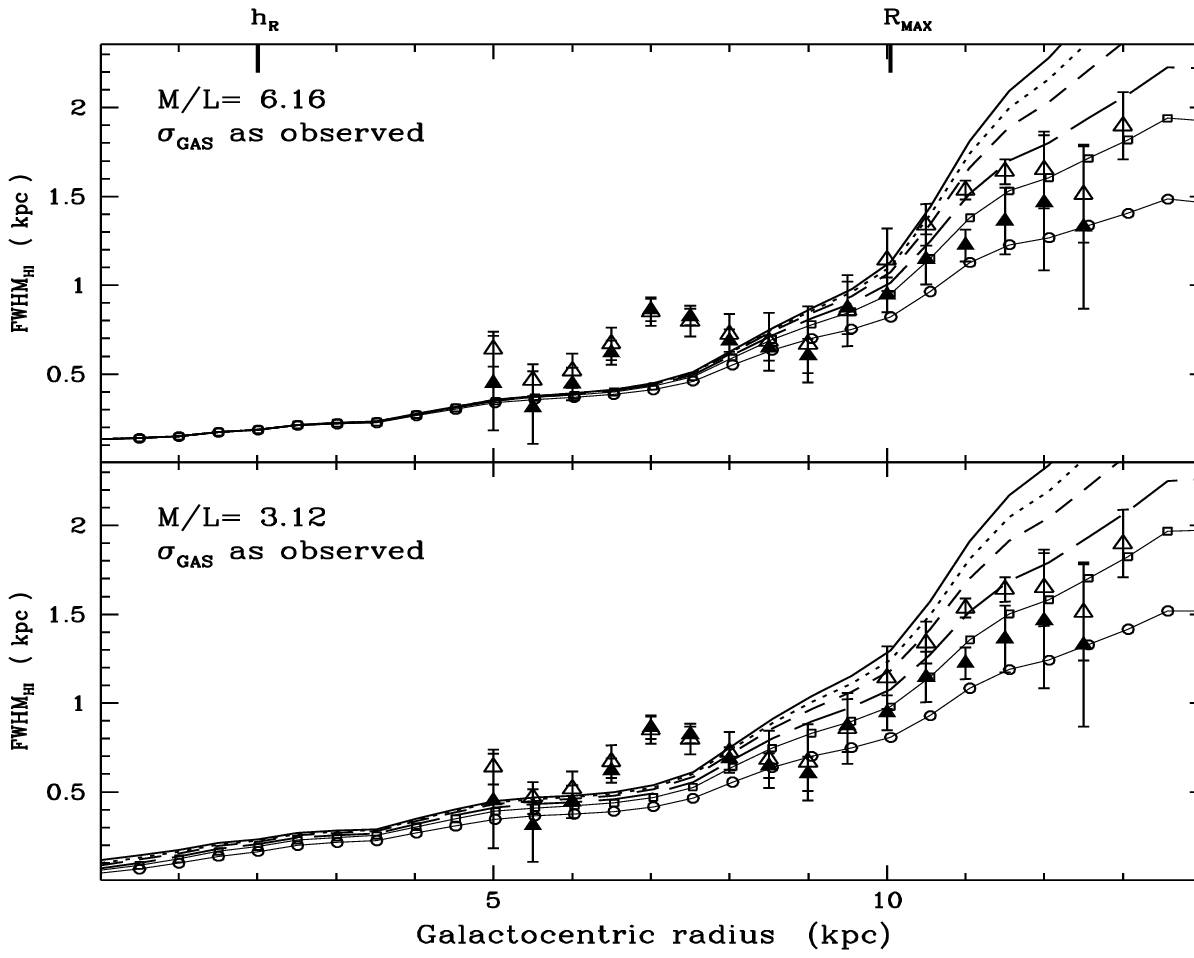,height=95mm,width=7.2cm}

\vspace*{-79mm} \noindent \parbox{4.5cm}{ Fig.  2.  The measured gas
layer widths (open and filled triangles for the i=84\add5 and
warp-included case, respectively).  The drawn lines correspond to models
with different halo shape:\qrho=1.0, 0.7, 0.5, 0.3, 0.2, and 0.1, from
top to bottom.  Using both inclination cases I find: \qrho =
0.2$_{-0.1}^{+0.3}$.  The mass-to-light ratio of the stellar disk is not
constrained by these flaring measurements. 
}

\vspace*{7mm}

\noindent important an even
{\em denser}, i.e.  {\em flatter}, DM halo would be required to have a
gas layer as thin as observed.  And c) If the vertical velocity
dispersion is smaller than the measured planar dispersion, the DM halo
would be rounder than inferred above.  There is no observational
evidence that such might be the case.  Furthermore, because the
interstellar medium is likely to be in the warm neutral phase due to the
low pressure\,\cite{pM93}, the short collision times ($\leq 10^5$ year)
preclude any anisotropy in the velocity dispersion tensor.  We conclude
that the DM halo of NGC 4244 is significantly flattened, with $\qrho =
0.2_{-0.1}^{+0.3}$. 

\subsection{An Alternative Explanation ?}

Pfenniger \etal$\!$\cite{PCM94} reviewed disk-like molecular hydrogen as
a dark matter candidate.  The fact that in many galaxies the shape of
the rotation curve due to the gas is similar to the observed rotation
curve\,\cite{Bos81,CCBV90} could then be explained if $\sim$ 6\% of
the gaseous surface density is in atomic form.  Since the self-gravity
of the gas layer beyond the optical disk strongly affects the
flaring (Eqn  2) I investigate whether the cold gas
hypothesis is consistent with NGC 4244's flaring curve.  In order for
the dark disk to have a thickness equal to the \HI layer, the {\em
vertical} velocity dispersion of the dark disk must be $\sim$ 20\%
larger than the dispersion in the \HI.  Furthermore, to avoid radial
instabilities, the {\em planar} velocity dispersion has to be 20\% -
100\% larger than the vertical dispersion: a dark disk requires an
anisotropic velocity dispersion tensor\,\cite{PaperIII}.

\section{Results: The Milky Way}

The flaring of the Milky Way's gas layer can be used to constrain the
shape of the dark matter halo as well.  The analysis is complicated
however because the uncertainties in the distance to the galactic center
(\RSUN), the rotation speed at the solar circle (\VSUN), and the
scale-length of the stellar mass (\hR).  Sackett\,\cite{pdS96b} reviewed
these values and found the following: \RSUN = 7.8 \pmt 0.7 kpc, \VSUN =
200 \pmt 20 \kms, and \hR = 3 \pmt 1 kpc.  Due to independent
constraints on the Oort constants\,\cite{KLB86} $A-B$ ($=\VSUN/\RSUN$)
and $A$, small values for \RSUN requires small values for \VSUN, and
vice-versa.  I calculated flaring curves for Milky Way models with a
small, intermediate and large value for \RSUN: (\RSUN,\VSUN)=(7.1,180),
(\RSUN,\VSUN)=(7.8,200), and (\RSUN,\VSUN)=(8.5,220).  These Milky Way
models include a stellar bulge and disk\,\cite{smK92} which is truncated
at around (\RSUN+4.5) kpc\,\cite{RCM92}.  Various authors reported on
the thickness\,\cite{Wetal90,DS91,mikM92,sanM94,sanM95} and column- and
volume densities\,\cite{bB88,BCAMT88,GCBTM87,Wetal90} of the \HI \& \Ht
layers of the Milky Way.  These observations were all scaled using a fit
to the rotation curve of the inner\,\cite{sanM95} and
outer\,\cite{mikM92} Milky Way, scaled separately for the three \RSUN \&
\VSUN combinations.  Since the models which predict the thickness of the
gas layer cannot handle a multi-component interstellar medium, the Milky
Way's flaring data must be compared with the model predictions beyond
the solar circle, where the \Ht column densities are small.  In the
model calculations I used the velocity dispersions of the \HI (9.2 \kms)
as determined in the inner galaxy\,\cite{sanM95}. 

Fig.  3 compares the observed thickness of the gas layer with model
predictions for various halo flattenings.  Only those model flaring
curves are plotted for which the column density in identified
stars\,\cite{KG89} (35 \pmt 5 \Msun pc\rtp{-2}) and the total column
density within 1.1 kpc from the plane\,\cite{KG91} (71 \pmt 6 \Msun
pc\rtp{-2}) are within 3-$\sigma$ of the nominal values.

\vspace*{-2mm}
\psfig{file=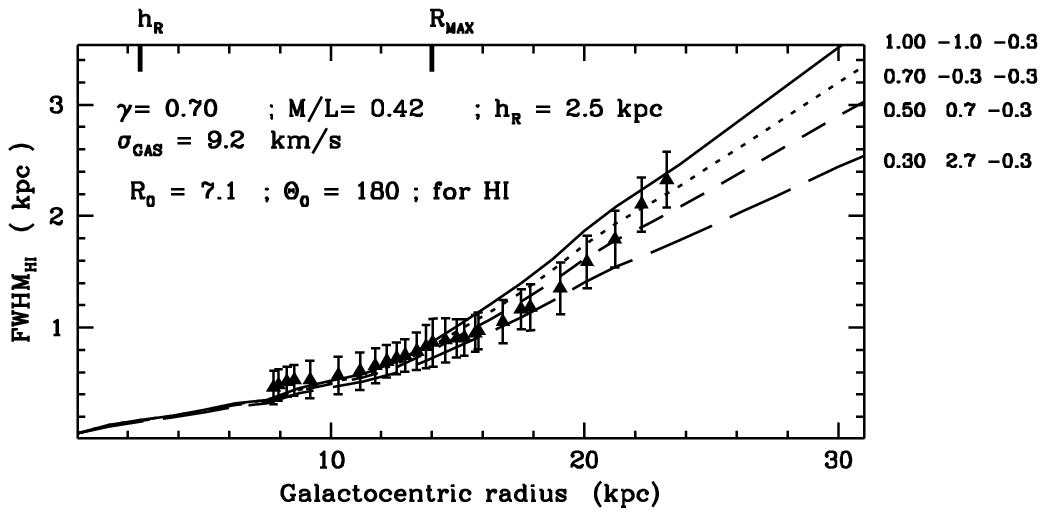,height=55mm,width=10cm}

\vspace*{-4mm} 
\begin{center} 

\noindent \parbox{10.5cm}{Fig.  3.  The flaring of the Milky Way
(triangles) superposed on model flaring curves with different halo
shapes.  For this assumed \RSUN-\VSUN combination, the rotation curve
has been determined till 18 kpc, and extrapolated beyond.  The last
three data points are thus least reliable.  This disk model is 70\%
($\gamma=0.70$) of maximal.  The errors are {\em not} random errors, but
are calculated by averaging the measurements of various authors (3 for
$R \le 18$ kpc, 2 beyond).  The three values plotted to the right of the
model curves indicate: a) the halo flattening, b) the difference between
the model and observed (stars+gas+halo) surface density density within
1.1 kpc, and c) the difference between the model stellar column density
and the column of identified stars (in units of their rms
uncertainties). 
}

\end{center}

\vspace*{2mm}

\noindent Model calculations with an optical scale-length of 2 kpc yield
a somewhat better correspondence between model and flaring data; longer
scale-lengths fit somewhat worse.  The same is true for Milky Way models
with larger values for \RSUN.  Because the errors on the observed widths
are systematic, no formal determination of the dark halo flattening can
be made.  For \RSUN=7.1 kpc, halo flattenings between 0.3 and 1.0 are
consistent with the observed flaring.  For larger values of \RSUN, the
correspondence between the model curves and observations become
progressively worse, while the inferred value for \qrho \ increases with
\RSUN as well: \qrho = 0.75 \pmt 0.25 for \RSUN=7.8 kpc.  For \VSUN =
220 \kms, the model flaring curves lie below the observations for all
choices of \RSUN, halo flattening and optical scale-length.

The inferred halo flattening depends only slightly upon the actual
values of the scale-length of the disk and its \MoverL{} because the
halo flattening influences the thickness of the gas layer only beyond
the optical disk\,\cite{PaperI}.  While mass models with \hR=3 kpc
require stellar disks which are (60 \pmt 10)\% of maximal, a disk with
an optical scale-length of 2 kpc requires $\gamma \sim 0.9 \pm 0.05$. 

For any given mass model, the midplane dark matter mass density at the
solar circle is found to be proportional to $q^{-0.72 \pm 0.02}$, with
an average value of: $\rho(z=0;q=1)$ = (9.4 \pmt 2) $10^{-3}$ \Msun
pc\rtp{-3} = (0.38 \pmt 0.08) protons cm\rtp{-3}.  Due to the strong
\qrho-dependence of \Rc(\qrho)\,\cite{PaperI}, the halo midplane density
depends more strongly upon \qrho \ near the center of the Galaxy than at
large radii (where $\rho_{DM}(z=0) \propto q\rtp{-0.5}$).

This determination of the flattening of the Milky Way's dark halo (0.65
\pmt 0.35) is consistent with previous
determinations\,\cite{BMO87,SZ90,vdM91,AC94} and insensitive to the
stellar mass distribution.  The major uncertainty is the value of the
gaseous velocity dispersion: a 15\% smaller dispersion would yield a
round halo.  These preliminary calculations favour Galaxy models with low
rotation speed and a small distance to the Galactic center. 

\section{Looking Ahead}

I have presented the results of a new method to determine the shape of
dark matter halos from sensitive \HI measurements and careful modeling. 
The first results exclude neither cold dark matter nor disk-like,
baryonic dark matter.  With current technology and ``reasonable''
observing times, the thickness of the \HI layer can be measured for
galaxies closer than $\sim$15 Mpc at inclinations $\ga 60^o$.  I
recently observed seven more systems (NGC 2366, 2403, 2903, 2841, 3521,
4236, and 5023) for which I will try to determine the DM halo shapes. 
Furthermore, the analysis of the flaring of the gas layers of M31 is in
progress.  With this increased sample it will be possible to gauge the
significance of the highly flattened halo of NGC 4244 and, hopefully,
put more stringent constraints on the nature of the dark matter.

\section*{Acknowledgements}

Most of the work presented here was part of my thesis project at
Columbia University.  I thank my advisor Jacqueline van Gorkom, Penny
Sackett for providing me with the disk-like surface density
distributions, and Mike Merrifield for suggestions to improve this
contribution.  This work was supported in part through an NSF grant
(AST-90-23254 to J.  van Gorkom) to Columbia University and PPARC grant
GR/K58227.  And of course I would like to thank Dr.  Ramachers and Prof. 
Klapdor-Kleingrothaus for a superbly organized conference and the
Max-Planck-Institut f\"{u}r Kernphysik conference for financial support.


\section{References}

\begin{thebibliography}{99}


\bibitem{ABBS85} Albada, T.S. van, Bahcall, J.N., Begeman, K. \& Sancisi, R.,
   \Jour{\Napj}{295}{305}{1985}
%\bibitem{AS86} Albada, T.S. van \& Sancisi, R., 
%   \Jour{Phil. Trans. R. Soc. Lond. A.}{320}{447}{1986}
\bibitem{AC94} Amendt, P. \& Cuddeford, P., 
   \Jour{\Napj}{435}{93}{1994}
%\bibitem{BLD95} Bahcall, N.A., Lubin, M.L. \& Dorman, V., 
%   \Jour{\Napjl}{447}{L81}{1995}
\bibitem{BH90} Bicay, M.D. \& Helou, G.,
   \Jour{\Napj}{362}{59}{1990}
\bibitem{BMO87} Binney, J., May, A. \& Ostriker, J.P., 
   \Jour{\Nmnras}{226}{149}{1987}
\bibitem{Bos81} Bosma, A., 
   \Jour{\Naj}{86}{1971}{1981}
%\bibitem{BHB92} Briel, U.G., Henry, J.P. \& Boehringer, H.,
%   \Jour{\Naap}{259}{L31}{1992}
%\bibitem{Bro95} Broeils, A.H.,
%   in {\em Dark Matter}, edited by S.S.  Holt, C.L.  Bennet (AIP conference
%   proceedings 336),  1995, p.  125
%\bibitem{Bro92} Broeils, A.H.,
%   Ph. D. Thesis, Rijksuniversiteit Groningen, 1992
\bibitem{BCAMT88} Bronfman, L., Cohen, R.S., Alvarez, H., May, J.,
   Thaddeus, P.,    \Jour{\Napj}{324}{248}{1988}
\bibitem{BC96a} Buote, D.A., Canizares, C.R.,
   \Jour{\Napj}{468}{184}{1996a}
%\bibitem{BC96b} Buote, D.A., Canizares, C.R.,
%    \Jour{\Napj}{457}{177}{1996b}
\bibitem{bB88} Burton, W.B., 1988, in {\em Galactic and Extragalactic
   Radio Astronomy}, edited by G.L. Verschuur and K.I. Kellermann
   (Springer-Verlag), Chapter 7
\bibitem{CCBV90} Carignan, C., Charbonneau, P., Boulanger, F. \& Viallefond, F.,
   \Jour{\Naap}{234}{43}{1990}
%\bibitem{CP90} Carignan, C. \& Puche, D., 
%   \Jour{\Naj}{100}{394}{1990}
\bibitem{CvG91} Casertano, S. \& van Gorkom, J. H.,
    \Jour{\Naj}{101}{1231}{1991}
%\bibitem{stC95} C\^{o}t\'{e}, S., Ph. D. Thesis, Australian National
%   University, (1995)
\bibitem{DS91} Diplas, A., Savage, D.S., \Jour{\Napj}{377}{126}{1991}
\bibitem{jD94} Dubinski, J., 
   \Jour{\Napj}{431}{617}{1994}
\bibitem{GCBTM87} Grabelsky, D.A., Cohen, R.S., Bronfman, L., 
   Thaddeus, P., May, J., \Jour{\Napj}{315}{122}{1987}
%\bibitem{GS95} Gerhard, O. \& Silk, J.,
%   astro-ph/9509149, submitted to {\Napj}, 1995
%\bibitem{GGST74} Gott, J.R., Gun, J.E., Schramm, D.N. \& Tinsley, B.M.,
%   \Jour{\Napj}{194}{543}{1974}
%\bibitem{jG80} Gunn, J.E., \Jour{Phil. Trans. R. Soc. Lond. A}{296}{313}{1980}
%\bibitem{HO86} Hegyi, D. \& Olive, K.A., 
%   \Jour{\Napj}{303}{56}{1986}
\bibitem{HS94} Hofner, P. \& Sparke, L.,
   \Jour{\Napj}{428}{466}{1994}
%\bibitem{nKjG91} Katz, N. \& Gunn, J.E.,
%   \Jour{\Napj}{377}{365}{1991}
\bibitem{smK92} Kent, S.M., 
   \Jour{\Napj}{387}{181}{1992}
\bibitem{KLB86} Kerr, F.J., Lynden-Bell, D., 
   \Jour{\Nmnras}{221}{1023}{1986}
%\bibitem{vdK81} Kruit, P.C. van der,
%   \Jour{\Naap}{99}{298}{1981}
\bibitem{KG91} Kuijken, K., Gilmore, G., 
   \Jour{\Napjl}{367}{L9}{1991}
\bibitem{KG89} Kuijken, K., Gilmore, G., 
   \Jour{\Nmnras}{239}{605}{1989}
%\bibitem{KBH82} Kulkarni, S.R., Blitz, L., Heiles, C., 
%   \Jour{\Napjl}{259}{L63}{1982}
\bibitem{LF89} Lake, G. \& Feinswog, L.,
   \Jour{\Naj}{98}{166}{1989}
%\bibitem{LG91} Lockman, F.J., Gehman, C.S., 
%   \Jour{\Napj}{382}{182}{1991}
\bibitem{sanM95} Malhotra, S., 
   \Jour{\Napj}{448}{138}{1995}
\bibitem{sanM94} Malhotra, S., 
   \Jour{\Napj}{443}{687}{1994}
\bibitem{pM93} Maloney, P., 
   \Jour{\Napj}{414}{41}{1993}
\bibitem{vdM91} Marel, R. P. van der,
   \Jour{\Nmnras}{248}{515}{1991}
\bibitem{mikM92} Merrifield, M.R., 
   \Jour{\Naj}{103}{1552}{1992}
%\bibitem{MLAF95} Mushotzky, R.F., Loewenstein, M., Arnaud, K. \& Fukazawa, Y 
%   in {\em Dark Matter}, edited by S.S.  Holt, C.L.  Bennet (AIP conference
%   proceedings 336), 1995, p.  231
\bibitem{NFW96} Navarro, J.F., Frenk, C.S., \& White, S.D.M,
    \Jour{\Napj}{462}{563}{1996}
\bibitem{PaperI}   Olling, R.P.,
   \Jour{\Naj}{110}{591-612}{1995,  [{\bf Paper   I}]}
\bibitem{PaperII}  Olling, R.P.,
   \Jour{\Naj}{112}{457-480}{1996a, [{\bf Paper  II}]}
\bibitem{PaperIII} Olling, R.P., 
   \Jour{\Naj}{112}{481-490}{1996b, [{\bf Paper III}]}
%\bibitem{OM96}  Olling, R.P., Merrifield, M.R., in preparation (1996)
\bibitem{PSS96} Persic, M., Salucci, P., Stel, F.,
    \Jour{\Nmnras}{281}{27}{1996}
\bibitem{PCM94} Pfenniger, D., Combes, F. \& Martinet, L.,
    \Jour{\Naap}{285}{79}{1994}
\bibitem{RCM92} Robin, A.C., Cr\'{e}z\'{e}, M., Mohan, V., 
   \Jour{\Napjl}{400}{L25}{19192}
%\bibitem{vcR93} Rubin, V.C., 
%   \Jour{Proc. Natl. Acad. Sci. USA}{90}{4814}{1993}
\bibitem{pdS96b} Sackett, P.D., submitted to \Napj, astro-ph/9608164 (1996b)
%\bibitem{pdS96} Sackett, P.D., ``The Distribution of Dark Mass in
%   Galaxies: Techniques, Puzzles, and Implications for Lensing,'' in
%   {\em Proceedings of IAU 173, Astrophysical Applications of Gravitational
%   Lensing,\/} eds.  C.  Kochanek and J.  Hewitt (Dordrecht: Kluwer), 1996, p. 
%   165-175, astro-ph/9508098
\bibitem{SP95} Sackett, P.D. \& Pogge R.W., in {\em Dark Matter}, 
   edited by S.S.  Holt, C.L.  Bennet (AIP conference proceedings
   336), 1995, p.  141
\bibitem{SRJF94} Sackett, P.D., Rix, H.W., Jarvis, B.J. \& Freeman, K,C.,
   \Jour{\Napj}{436}{629}{1994}
\bibitem{SS90} Sackett, P.D. \& Sparke, L.S.,
   \Jour{\Napj}{361}{408}{1990}
\bibitem{SZ90} Sommer-Larsen, J. \& Zhen, C., 
   \Jour{\Nmnras}{242}{10}{1990}
\bibitem{SKD92} Steiman-Cameron, T.Y., Kormendy, J. \& Durisen, R.H.,
  \Jour{\Naj}{104}{1339}{1992}
%\bibitem{UM94} Udry, S. \& Martinet, L.,
%   \Jour{\Naap}{281}{314}{1994}
%\bibitem{VWRH95} Vogel, S.N., Weymann, R., Rauch, M. \& Hamilton, T.,
%   \Jour{\Napj}{441}{162}{1995}
%\bibitem{WSSOK91} Walker, T.P., Steigman, G., Schramm, D.M., Olive, K.A. \&
%   Kang, H.,  \Jour{\Napj}{376}{51}{1991}
\bibitem{Wetal90} Wouterloot, J.G.A., Brand, J., Burton, W.B., Kwee, K.K.,
   \Jour{\Naap}{230}{21}{1990}
\end{thebibliography}
\end{document}